\documentclass{an}
\usepackage{graphicx}
\usepackage{times}
\usepackage{fancyhdr}
\newcommand{\beq}{\begin{equation}}
\newcommand{\eeq}{\end{equation}}
\newcommand{\bea}{\begin{eqnarray}}
\newcommand{\eea}{\end{eqnarray}}

\def\laq{~\raise 0.4ex\hbox{$<$}\kern -0.8em\lower 0.62
ex\hbox{$\sim$}~}
\def\gaq{~\raise 0.4ex\hbox{$>$}\kern -0.7em\lower 0.62
ex\hbox{$\sim$}~}

\def \pa {\partial}

\def \b {\beta}
\def \a {\alpha}

\def \ga {\gamma}

\def \ep {\epsilon}

\def \om {\omega}
\def \Om {\Omega}
\def \noi {\noindent}

\begin{document}

\begin{titlepage}

\thispagestyle{empty}

\begin{flushright}
CERN-PH-TH/2005-160\\
astro-ph/0509174
\end{flushright}

\vspace{2 cm}

\begin{center}
\Large\bf Primordial magnetic seeds from string cosmology
\end{center}

\vspace{1.cm}

\begin{center}
M. Gasperini$^{1,2,3}$ 

\bigskip
$^{1}${\em CERN, Theory Unit, Physics Department, CH-1211 Geneva 23, Switzerland} 

\smallskip
$^{2}${\em Dipartimento di Fisica, Universit\`a di Bari, 
Via G. Amendola 173, 70126 Bari, Italy}

\smallskip
$^{3}${\em Istituto Nazionale di Fisica Nucleare, Sezione di Bari, 
Bari, Italy}

\end{center}

\vspace{1.5cm}

\centerline{\bf Abstract}

\bigskip
\noi
\baselineskip=15pt
After a discussion of the inflationary production of primordial magnetic seeds, and a short review of various possible mechanisms, we concentrate on the analysis of the photon--dilaton coupling typical of string theory models. Particular attention is paid to the constraints to be imposed on the primordial seed spectrum, and to the possibility of obtaining phenomenological signatures of heterotic and Type I superstrings, in principle accessible to present (or near-future) observations.

\vspace{5cm}
\begin{center}

To appear in the Proc. of the Int. Conference on \\ 
\smallskip
{\em ``The Origin and Evolution of Cosmic Magnetism"}, 
Bologna, 29 August -- 2 September 2005, \\ 
eds. R. Beck, G. Brunetti, L. Feretti and B. Gaensler 
({\em Atronomische Nachrichten}, Wiley, 2005)
\end{center}

\vspace{1.5cm}
\vfill
\begin{flushleft}
CERN-PH-TH/2005-160\\
September 2005 
\end{flushleft}

\end{titlepage}

\thispagestyle{empty}
\vbox{}
\newpage

\title{Primordial magnetic seeds from string cosmology}

\author{M. Gasperini\inst{1,2,3}}
\institute{CERN, Theory Unit, Physics Department, CH-1211 Geneva 23, Switzerland
\and
Dipartimento di Fisica, Universit\`a di Bari, 
Via G. Amendola 173, 70126 Bari, Italy\thanks{Permanent address.} 
\and
Istituto Nazionale di Fisica Nucleare, Sezione di Bari, 
Bari, Italy}

\date{Received ...; accepted ...; published online ...;}

\abstract{After a discussion of the inflationary production of primordial magnetic seeds, and a short review of various possible mechanisms, we concentrate on the analysis of the photon--dilaton coupling typical of string theory models. Particular attention is paid to the constraints to be imposed on the primordial seed spectrum, and to the possibility of obtaining phenomenological signatures of heterotic and Type I superstrings, in principle accessible to present (or near-future) observations. 
}

\correspondence{gasperini@ba.infn.it}

\maketitle

\section{Introduction}
Primordial magnetic seeds, produced during the very early stages of the cosmological evolution (possibly as the result of fundamental, high-energy interactions), are a recognized crucial ingredient for the generation of large-scale magnetic fields (see e.g.  Grasso \& Rubinstein (2001), Giovannini (2004)). How can they be produced? 

The most natural mechanism of primordial-seed production is, at least in principle, inflation.  Even in vacuum, indeed, the electromagnetic field $F_{\mu\nu}$ is non-zero because of its quantum fluctuations. During inflation these tend to be frozen outside the horizon, and they are eventually amplified when they re-enter inside the horizon, after the end of the accelerated epoch. Thus, inflation should produce a primordial spectrum of large-scale electromagnetic fields, just like the inflationary amplification of the metric fluctuations produces {\em scalar seeds} for the CMB anisotropies. 

Unfortunately, however, such a mechanism does not work when the electromagnetic field is minimally coupled to a conformally invariant geometry, represented by the metric 
\beq
ds^2= a^2(\eta)(d\eta^2 -dx_i^2). 
\eeq
The reason is that, because of the conformal invariance of the four-dimensional Maxwell Lagrangian, $\sqrt{-g}\,  g^{\mu\a}g^{\nu\b}F_{\mu\nu}F_{\a\b}$, the canonical action for the quantum fluctuations turns out to be completely decoupled from the geometry. If we write such an action (for instance in the radiation gauge $A_0=0=\pa_iA_i$), we obtain indeed the quadratic form
\beq
S=(1/2)\int d\eta\left(A_i^{\prime 2}+A_i \nabla^2 A_i\right)  
\eeq
(the prime denotes the derivative with respect to the conformal time $\eta$), 
and we are lead to free, oscillating, vacuum solutions for the electromagnetic (e.m.) fluctuations, 
\beq
A_i^{\prime \prime}+k^2 A_i=0, ~~ \Rightarrow ~~ 
A_i = {e^{-ik\eta}\over \sqrt{2k}},
\eeq
which are completely insensitive to the cosmological (even inflationary) expansion. So, there is no amplification of the fluctuations, and no production of e.m. seeds at all. 

How can we avoid this conclusion? There are various possibilities, at least in principle. We may assume, for instance, that:  
\begin{itemize}
\item[1.] the geometry is not exactly conformally flat, or that 
\item[2.] the coupling of the e.m. fields to the geometry is not exactly conformally invariant, or that 
\item[3.] there are non-conformal couplings to other background fields. 
\end{itemize} 
The string-cosmology mechanism belongs to the third point of this list. Before discussing it, it may be instructive to present a short review of the various (not necessarily alternative) attempts at breaking the conformal symmetry of the e.m. equations, so as to obtain a significant cosmological production of primordial magnetic seeds. 

\section{Inflationary seed production}
The first (and simplest) possibility of the above list is to modify the geometry  {\em without} modifying the Maxwell action, assuming for instance that the metric describes a Kasner-like, anisotropic expansion (Lotze 1990).  However, this leads  to an anisotropic seed production, which is strongly constrained by the present bounds on large scale anisotropy. 

Another (probably more realistic) example can be obtained by considering a higher-dimensional space-time manifold which is the product of $3$ {\em external} and $n$ {\em internal} (dynamical) dimensions, with scale factor $a$ and $b$ respectively, described by the metric: 
\beq
ds^2= a^2(\eta)(d\eta^2 -dx_i^2) - b^2(\eta) dy_a^2.  
\eeq
The corresponding Maxwell action contains the coupling to the internal geometry, 
\beq
S=(1/2)\int d\eta\, b^n\left(A_i^{\prime 2}+A_i \nabla^2 A_i\right), 
\eeq
and the internal scale factor $b$ plays the role of a time-dependent  ``pump field" for the amplification of the (quantum) canonical fluctuations $\psi_i=b^{n/2}A_i$, which satisfy the evolution equation
\beq
\psi_i^{\prime \prime} + k^2 \psi_i-b^{-n/2}(b^{n/2})^{\prime \prime} \psi_i=0.
\eeq
It is possible to produce magnetic seeds (Giovannini 2000), but this seems to require an expanding internal geometry, which may look not so natural. 

If we want to keep instead a four-dimensional, conformally flat metric (which seems to be preferred by inflation, after all), we may resort to the second possibility of the previous list,  breaking explicitly the conformal invariance of the Maxwell action. For instance, by adding  direct, non-minimal couplings of the photon field to the geometric curvature, such as $RF_{\mu\nu}F^{\mu\nu}$, or $R^{\mu\nu}A_\mu A_\nu$, or $RA_\mu A^\mu$, and so on  (Turner \& Widrow 1988, Basset et al. 2000). Terms like the last two, containing explicitly the vector $A_\mu$, break however the e.m. gauge invariance, and have to be strongly suppressed. We may  also consider the breaking of conformal symmetry induced at the quantum level by the so-called ``trace anomaly"  of the e.m. stress tensor (Dolgov 1993). In both cases there are, however, strong phenomenological constraints, and the allowed amplification of the e.m. fluctuations turns out to be small in a realistic model. 

I should also mention a very recent attempt, based on the addition to the Maxwell action  of non-conformally invariant boundary terms such as  $M^{-2}\int \nabla_\mu (A^2 \nabla^2 K \xi^\mu)$ ($K$ is the extrinsic curvature and $\xi$ the conformal Killing vector), 
which may acquire non-trival dynamical effects in the context of an appropriate ``trans-Planckian" model of  high-energy  gravitational interactions (Ashoorioon \& Mann 2005). The efficiency of such a breaking term, where seed production is concerned, seems to require, however, an effective mass $M$, which is about ten times smaller than the Hubble mass scale, $M \sim 10^{-34}$ eV, and which corresponds to an enormous (and probably not very realistic, in my opinion) coupling constant $M^{-2}$. 

The third, more conservative point of the previous list is based on the introduction of non-conformal-invariant couplings of the photon to some background field other than the metric. For instance: 
\begin{itemize}
\item  non-minimal coupling to the inflaton (Ratra 1992);
\item  non-minimal coupling to the axion (Garretson, Field \& Carrol  1992; Brandenberger \& Zhang 1999); 
\item coupling to the dilaton (Gasperini, Giovannini \& Veneziano 1995;  Lemoine \& Lemoine 1995);
\item minimal coupling to charged scalar fields (Calzetta, Kandus \& Mazzitelli, 1998; Giovannini \& Shaposhnikov 2000; Davis, Dimopoulos \& Prokopec 2001; Finelli \& Gruppuso 2001);
\item coupling to a vector field that breaks spontaneously the Lorentz symmetry (Bertolami \& Mota 1999);
\item coupling to the metric and matter perturbations (Marklund, Dunsby \& Brodin 2000; Tsagas,  Dunsby \& Marklund  2003; Maroto 2001; Matarrese et al. 2005; Takahashi et al. 2005); 
\item coupling to graviphotons, which are massive vector fields required by supersymmetry (Gasperini 2001).
\end{itemize}

This list is certainly incomplete, and I must apologize for the missing references. I hope, nevertheless, that the number of reported attempts may give a clear feeling of the importance of finding an efficient and realistic mechanism for the production of primordial magnetic seeds. 
The mechanism that is possibly in action in the string-cosmology context will be presented, with some details, in the next section. 

\section{Photon-dilaton interaction in string cosmology}

The string effective action, which in principle describes the low-energy dynamics of all background fields, contains in general a direct coupling between the scalar dilaton field and the e.m. field that appears 
even at tree-level in the loop expansion, and to lowest order in the $\a'$ (higher-derivative) corrections: 
\beq
S=-(1/4)\int d^4x \sqrt{-g}\, e^{-\ep \phi} F_{\mu\nu}F^{\mu\nu}.
\eeq
Here $\phi$ represents the dimensionally reduced dilaton, which includes the proper volume of the internal dimensions (taking into account their possible dynamical contribution), and $\ep$ is a constant parameter that depends on the given string theory model. For instance, $\ep=1$ for the heterotic superstring,  and $\ep=1/2$ for the Type I superstring (see e.g. Lidsey, Wands \& Copeland (2000)). For the above action, the canonical e.m. fluctuations $\psi_i=e^{-\ep \phi/2} A_i$ are coupled to the dilaton field, and are possibly amplified according to the evolution equation
\beq
\psi_i^{\prime \prime} + k^2 \psi_i-e^{\ep \phi/2}(e^{-\ep \phi/2})^{\prime \prime} \psi_i=0.
\eeq

It should be stressed that the main difference between this mechanism and other mechanisms of e.m. amplification is that the coupling to the dilaton appearing in the above equation has not been invented {\em ad hoc}, just to produce seed fields, but is rigidly prescribed by the chosen string theory model. As a consequence, the presence of this coupling does not necessarily guarantee an efficient seed production, since the  amplification process strongly depends on various details of the dilaton evolution. This suggests an interesting  possibility: the existing phenomenological constraints on the process of seed production could  be used as constraints on the various string theory models of the early Universe, and on the parameters of the string effective action. 

\begin{figure}
\resizebox{\hsize}{!}
{\includegraphics[]{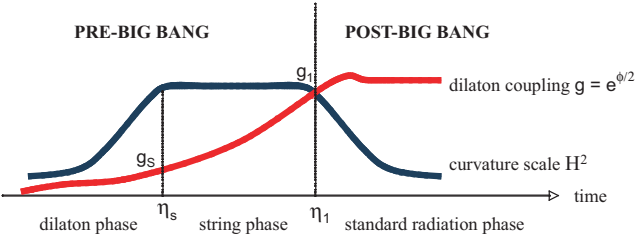}}
\caption{Time evolution of the exponentiated dilaton, $g=\exp \phi$, and of the curvature scale $H^2$, in a typical example of ``minimal" string-cosmology scenario. The standard, radiation-dominated phase is preceded by a high-curvature string phase (extending from $\eta_s$ to $\eta_1$), and by a low-energy dilaton phase where both $g$ and  $H^2$ tend to zero as the time coordinate approaches $-\infty$.}
\label{f1}
\end{figure}

In order to illustrate this possibility, I will consider here one of the simplest, typical examples of string-cosmology model: the so-called ``minimal" pre-big bang scenario (see e.g. Gasperini \& Veneziano (2003)), illustrated in Fig. \ref{f1}. The curvature scale $H^2$ (the blue curve) grows in time during the dilaton-driven phase, reaches a maximum fixed by the string scale, and then decreases, following the usual behaviour of the standard (radiation-dominated) cosmoloy. The dilaton field $g=\exp(\phi/2)$ (the red curve) keeps growing during the whole pre-big bang phase, and tends to be frozen in the subsequent phase of standard evolution. The primordial, pre-big bang regime includes an initial, dilaton-dominated phase whose dynamics is fully determined by the low-energy string effective action (Veneziano 1991; Gasperini \& Veneziano 1993), and a subsequent, high-curvature string phase (Gasperini et al. 1995; Brustein et al. 1995), characterized by two unknown parameters: the duration 
$z_s=\eta_s/\eta_1$ (from the initial time $\eta_s$ to the final time $\eta_1$), and the acceleration, which we may express as the ratio between the initial and final values of the dilaton, $g_s/g_1$. 

The computation of the amplified e.m. perturbations, for this class of models, leads to a two-parameter spectrum. Considering for the moment the heterotic model characterized by $\ep=1$, we obtain that the spectral energy density of the produced  seeds, in critical units, can be written as (Gasperini et al. 1995):  
\bea
&&
\Om_\ga(\om) \simeq \Om_r(t_0)\left(M_{\rm s}\over M_{\rm P}\right)^2\left(\om \over \om_1\right)^{3-2\nu},   ~~~~~~~\om_s< \om <\om_1
\nonumber\\
&&
 \simeq \Om_r(t_0)\left(M_{\rm s}\over M_{\rm P}\right)^2\left(\om_s \over \om_1\right)^{3-2\nu} \left(\om\over \om_s\right)^{4-\sqrt 3},
~~~~\om<\om_s, 
\nonumber\\ &&
\label{9}
\eea
where $2\nu= |1-2\b|$, and $\b$ is the constant power that controls the time evolution of the dilaton, $ \exp (-\phi/2) \sim |\eta|^\b$.   
The two unknown parameters are the slope ($3-2\nu$) and the extension ($\om_s/\om_1$) of the high-frequency branch of the spectrum. The end point of the spectrum, however, is fixed, and is fully controlled by the fundamental ratio between string and Planck mass, $M_{\rm s}/M_{\rm P}$ (which is expected to be a number of order $0.1$, according to unified models of all interactions). It turns out, in particular, that the present value of the cut-off frequency is in the range   $\om_1\sim (M_{\rm s}/M_{\rm P})^{1/2}\,10^{11}$ Hz, and 
the present peak value of the spectrum is $\Om_\ga(\om_1)\sim 
(M_{\rm s}/M_{\rm P})^{2}\Om_r(t_0)$, where $\Om_r(t_0)$ 
is the present critical fraction of radiation energy density. 

We are now in the position of considering the main constraints on the amplified seed spectrum. A first, model-independent constraint is that the e.m. fluctuations should not destroy the large-scale homogeneity of the cosmological background, and should be treated as small perturbations with negligible back-reaction. This imposes the stringent constraint
\beq
\Om_\ga(\om, t)/\Om_r(t)<1,
\label{10}
\eeq
to be satisfied at all times, and all frequency scales. 

A second, important constraint is that the amplification should be efficient enough to seed the cosmic galactic fields. This condition, however, is model-dependent, and just to give an idea I will present here a {\em naive} example assuming the existence of a standard galactic dynamo, operating since the epoch of galaxy formation. The required seed field is then $\sim10^{-19}$ Gauss, and the required energy density, assuming flux conservation, can be expressed in critical units as follows (Turner \& Widrow 1988):
\beq
\Om_\ga(\om_G, t_0)/\Om_r(t_0) \gaq 10^{-34}.
\label{11}
\eeq
The required coherence scale corresponds to a frequency scale that is roughly $10^4$ larger than the scale of the present Hubble horizon,
$\om_G(t_0) \sim (1\, {\rm Mpc})^{-1} \sim 10^{-14}$ Hz. 

\begin{figure}
\resizebox{\hsize}{!}
{\includegraphics[]{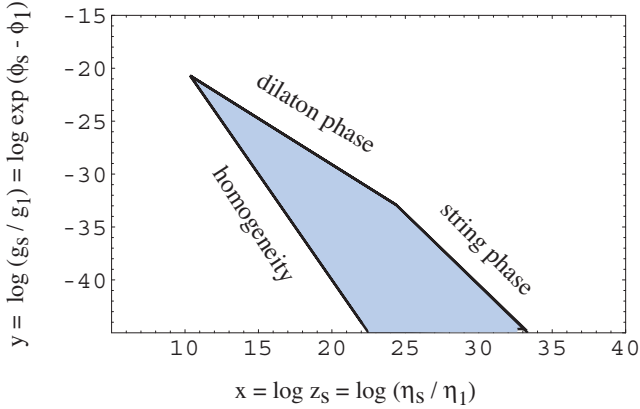}}
\caption{Allowed region for efficient and consistent seed production, in the two-dimensional parameter space of minimal pre-big bang 
models.}
\label{f2}
\end{figure}

The  above two conditions can be well satisfied if we consider a minimal pre-big bang scenario based  on the heterotic superstring model, and if we assume a realistic value of the fundamental string to Planck mass ratio not very different from the typical value $M_{\rm s}/M_{\rm P} \sim 0.1$. The allowed region of the parameter space, satisfying the conditions  (\ref{10}), (\ref{11}) imposed on the spectrum (\ref{9}), is illustrated in Fig. \ref{f2}, where the left border line corresponds to the homogeneity condition, the upper-right border line to the  condition for efficient seed production in the dilaton phase, and the lower-right border line to the  condition for efficient seed production in the string phase. The mechanism is clearly successful, provided $z_s \gaq 10^{10}$ and $g_s/g_1\laq 10^{-20}$, i.e. provided there is a {\em large} enough and {\em fast} enough growth of the dilaton during the string phase.

It may be important to stress that there are in principle independent checks  of such a mechanism of seed production, based on the fact that the amplification of electromagnetic fluctuations is always associated with the amplifications of tensor metric perturbations, and to the consequent formation of a cosmic background of relic gravitons. 
The graviton spectrum is closely related to the electromagnetic  spectrum, and the  phenomenological constraints determining the allowed region for an efficient seed mechanism can thus directly affect the shape of the graviton spectrum. For the heterotic model, in particular, it turns out that there is no overlap between the region of parameter space corresponding to an efficient seed production, and the region corresponding to the formation of background of cosmic gravitons detectable by present (and advanced) generations of gravitational detectors (Gasperini 1996). 

This conclusion may be evaded, however, in the context of string models with a photon--dilaton coupling parameter $\ep \not=1$. In that case the spectrum (\ref{9}) is modified as follows,
\bea
&&
\Om_\ga(\om) \simeq \Om_r(t_0)\left(M_{\rm s}\over M_{\rm P}\right)^2\left(\om \over \om_1\right)^{3-2\mu},   ~~~~~\om_s< \om <\om_1
\nonumber\\
&&
 \simeq \Om_r(t_0)\left(M_{\rm s}\over M_{\rm P}\right)^2\left(\om_s \over \om_1\right)^{3-2\mu} \left(\om\over \om_s\right)^{3-|1 -\ep\sqrt 3|},
\om<\om_s, 
\nonumber\\ &&
\label{12}
\eea
with $2\mu= |1-2\ep\b|$, and the overlap between the allowed regions for photons and gravitons becomes possible, in principle. In any case, one can obtain independent tests of the considered string cosmology scenario. 

Let us suppose, for instance, that the heterotic superstring model is the correct unified scheme for the description of our primordial Universe; let us also suppose that the next generation of gravitational antennas, such as LIGO or VIRGO, will detect a cosmic background of relic gravitons, fitting the minimal string-cosmology predictions: one must then deduce that there is no room, in such a context, for an efficient production of primordial magnetic seeds. 

Vice versa, let us suppose that  the cosmic gravitons will be detected, and that we will find  some independent and direct confirmation that there is also the expected seed production: this would imply that the heterotic model is not the appropriate theoretical tool for the construction of a realistic cosmological scenario, and that models with $\ep\not=1$ are favoured instead. Work along these lines is currently  in progress. 

\section{Conclusion}
The inflationary production of primordial magnetic seeds is an interesting open problem of contemporary astrophysics. String theory seems to suggest a possible solution based on the cosmological evolution of the dilaton, and on its coupling to the electromagnetic field. The consistency of this solution imposes non-trivial constraints  and thus provides important (direct and indirect) information on the primordial history of our Universe, and on Planck/string scale physics and cosmology. 

\acknowledgements
I greatly appreciated the warm hospitality of the staff of the Institute of Radioastronomy of Bologna, and their perfect organization of this stimulating and successful Conference. I am also grateful to Massimo Giovannini for many clarifying discussions, and for his precious help in the preparation of this contribution.

\end{document}